\documentclass[aps,nofootinbib,twocolumn,groupedaddress,floatfix,eqsecnum]{revtex4}
\usepackage{graphicx}
\usepackage{epsfig}
\usepackage{epstopdf}
\usepackage[mathscr]{eucal}
\usepackage{amsmath,amssymb,bm,graphics}
\usepackage[pdftex,letterpaper=true]{hyperref}

\newcommand{\half}{{\frac{1}{2}}}
\newcommand{\ket}[1]{\,\left|\,{#1}\right\rangle}

\newcommand{\mbf}[1]{\mathbf{#1}}

\def\Dslash{\raise.15ex\hbox{/}\kern-.7em D}
\def\Pslash{\raise.15ex\hbox{/}\kern-.7em P}

\usepackage{color}

\renewcommand{\bar}[1]{\overline{#1}}
\begin{document}

\preprint{SLAC-PUB-13192}

\title{Light-Front Dynamics and AdS/QCD Correspondence: \\
Gravitational Form Factors of Composite Hadrons}

\author{Stanley J. Brodsky}
%\email[]{sjbth@slac.stanford.edu}
%\homepage[]{Your web page}
%\thanks{}
%\altaffiliation{}
\affiliation{Stanford Linear Accelerator Center, Stanford University,
Stanford, California 94309, USA}

\author{Guy F. de T\'eramond}
%\email[]{gdt@asterix.crnet.cr}
%\homepage[]{Your web page}
%\thanks{}
%\altaffiliation{}
\affiliation{Centre de Physique Th\'eorique,  Ecole Polytechnique, 91128 Palaiseau, France and \\
Universidad de Costa Rica, San Jos\'e, Costa Rica}

\date{\today}

\begin{abstract}

Light-front holography is a remarkable feature of the AdS/CFT correspondence between gravity in AdS space and conformal field theories in 
physical space-time; it allows string modes  $\Phi(z)$ in the anti-de Sitter (AdS) fifth dimension
to be precisely mapped to the light-front wavefunctions of hadrons in physical space-time in
terms of a specific light-front impact variable $\zeta$ which measures the separation of the
quark and gluonic  constituents  within the hadron.
This mapping was originally obtained by matching the exact expression for electromagnetic current matrix elements in AdS space with the corresponding exact expression for the current matrix element using light-front theory in physical space-time.  In this paper we show that one obtains the identical holographic mapping using matrix elements of the energy-momentum tensor.  To prove this, we show that there exists a correspondence between the matrix elements of the energy-momentum
tensor  of the fundamental hadronic constituents in QCD with the transition amplitudes
describing the interaction of string modes in
AdS space with an external graviton field which propagates in the AdS interior. The agreement of the results for electromagnetic and gravitational hadronic transition amplitudes provides an important consistency test and verification of holographic mapping from AdS to physical observables defined on the light-front.

\end{abstract}

\pacs{11.15.Tk, 11.25.Tq, 12.38.Aw, 12.40.Yx}

\maketitle

\section{Introduction}

One of the most challenging problems of strong interaction dynamics is to determine the composition of hadrons in terms of their fundamental QCD quark and gluon degrees of freedom. Because of the 
strongly coupling nature of QCD in the infrared domain, it has been difficult to find analytic solutions for the wavefunctions of hadrons or to make precise predictions for hadronic properties outside of the perturbative regime.  Thus an important theoretical goal is to find an initial approximation to bound state problems in QCD which is analytically tractable and which can be systematically improved.
Recently 
the AdS/CFT correspondence~\cite{Maldacena:1997re} between string
states in anti--de Sitter (AdS) space and conformal field theories (CFT) in physical space-time, modified for color confinement,
has led to a semiclassical model for strongly-coupled QCD which provides analytical insights
into its inherently non-perturbative nature including hadronic spectra, decay constants, and wavefunctions.

As we have shown recently, there is a remarkable mapping between the AdS description of hadrons and the Hamiltonian formulation of QCD in physical space-time quantized on the light front. 
The light-front wavefunctions of bound states in QCD are relativistic and frame-independent
generalizations of the familiar Schr\"odinger wavefunctions of
atomic physics, but they are determined at fixed light-cone time
$\tau  = t +z/c$---the ``front form" advocated by Dirac~\cite{Dirac:1949cp}---rather than
at fixed ordinary time $t$.  
The  light-front wavefunctions  of a
hadron are independent of the momentum of the hadron, and they are
thus boost invariant; Wigner transformations and Melosh rotations
are not required. The light-front formalism for gauge theories in
light-cone gauge is particularly useful in that there are no ghosts,
and one has a direct physical interpretation of  orbital angular
momentum.

{\it Light-Front Holography} is an important feature of AdS/CFT; it allows string modes $\Phi(z)$ in the AdS fifth dimension
to be precisely mapped to the light-front wavefunctions of hadrons in physical space-time in
terms of a specific light-front impact variable $\zeta$ which measures the separation of the
quark and gluonic  constituents  within the hadron.
This mapping was originally obtained by matching the exact expression for electromagnetic current matrix elements in AdS space with the corresponding exact expression for the current matrix element using light-front theory in physical space-time~\cite{Brodsky:2006uqa, Brodsky:2007hb}. In this paper we shall show that one obtains the identical holographic mapping using the matrix elements of the energy-momentum tensor.  To prove this new result, we will show that there exists a correspondence between the matrix elements of the energy-momentum tensor  of the 
fundamental hadronic constituents in QCD with the transition amplitudes
describing the interaction of string modes in
anti-de Sitter space with an external graviton field which propagates in the AdS interior. 

The AdS/CFT correspondence implies
that a strongly-coupled gauge theory is equivalent to the propagation
of weakly-coupled strings in a higher dimensional space,
where physical quantities are computed in terms of an effective
gravitational theory. Thus, the AdS/CFT duality provides a gravity
description in a ($d+1$)-dimensional AdS
space-time in terms of a
$d$-dimensional conformally-invariant quantum field theory at the AdS 
asymptotic boundary~\cite{Gubser:1998bc, Witten:1998qj}.

Holographic duality requires one to consider a higher dimensional warped space with
negative curvature and a four-dimensional boundary.  In particular, the conformal isometries of the
five-dimensional anti-de Sitter space, a maximally symmetric space-time geometry
with negative curvature,  provides the basis for establishing a duality between a gravity or string theory on AdS$_5$ space and a conformal gauge theory defined at its four-dimensional space-time 
boundary.  In its original formulation~\cite{Maldacena:1997re}, a correspondence was
established between the supergravity approximation to type IIB superstring theory
on a curved background asymptotic to the product space of AdS$_5 \times S^5$~\cite{Schwarz:1983qr} 
and the large $N_C$, $\mathcal{N} = 4$,
supersymmetric Yang-Mills (SYM) gauge theory in four dimensions with gauge group
$SU(N)$~\cite{Brink:1976bc}. 
The group of conformal transformations
$SO(4,2)$ which acts at the asymptotic boundary of AdS space, acts also as 
the group of isometries of AdS$_5$, 
and $S^5$ corresponds to the $SU(4) \sim SO(6)$ global symmetry which 
rotates the particles
present in the SYM supermultiplet. The supergravity duality requires
a large AdS radius $R$ corresponding to a large value of the 't Hooft
parameter $g_s N_C$, where $R = ({4 \pi g_s N_C})^{1/4} \alpha_s'^{1/2}$ and
$\alpha_s'^{1/2}$ is the string scale. The classical  approximation
corresponds to the stiff limit where the
string tension $T = R^2 / 2 \pi \alpha' \to \infty$, effectively
suppressing string fluctuations.

QCD is fundamentally different from SYM theories where all the
matter fields transform in adjoint multiplets of $SU(N_C)$.
QCD is also a confining theory in the infrared
with a mass gap $\Lambda_{\rm QCD}$ and a well-defined spectrum of color singlet states.
Conformal symmetry is broken in physical QCD by quantum effects and quark masses.
There are indications however,  both from
theory and phenomenology, that the QCD coupling is slowly varying at small momentum 
transfer~\cite{Brodsky:2008pg}. In particular, a new extraction of the effective strong coupling constant
$\alpha_s^{g_1}(Q^2)$ from CEBAF Large Acceptance Spectrometer  Collaboration (CLAS) spin structure function data 
using the Bjorken sum $\Gamma_1^{p-n}(Q^2)$ 
in an extended $Q^2$ region~\cite{Deur:2008rf},
indicates the lack of $Q^2$ dependence of $\alpha_s$ in the low $Q^2$ limit. One can understand this
physically~\cite{Brodsky:2008pg}: in a confining theory where gluons have an effective 
mass or maximal wavelength, all vacuum polarization
corrections to the gluon self-energy decouple at long wavelength; thus an infrared fixed point appears to be a natural consequence of confinement~\cite{Cornwall:1981zr}. Furthermore, if one considers a
semiclassical approximation to QCD with massless quarks and without
particle creation or absorption, then the resulting $\beta$ function
is zero, the coupling is constant, and the approximate theory is
scale and conformal invariant~\cite{Parisi:1972zy}. One can use conformal symmetry as 
a {\it template}, systematically correcting for its nonzero $\beta$ function as well
as higher-twist effects~\cite{Brodsky:1985ve}.

Different values of the holographic variable $z$ determine the scale of the invariant
separation between the partonic constituents. 
Hard scattering processes occur in the small-$z$ ultraviolet (UV)
region of AdS space. In particular,
the $Q \to \infty$ zero separation limit corresponds to the $z \to 0$ asymptotic boundary, where the QCD
Lagrangian is defined. 
In the large-$z$ infrared (IR) region  a cut-off is introduced to truncate the regime where the AdS modes can propagate. The infrared cut-off breaks conformal invariance, allows the introduction of
a scale and a spectrum of particle states. In the hard-wall model~\cite{Polchinski:2001tt}
 a cut-off is placed at a
finite value $z_0 = 1/\Lambda_{\rm QCD}$ and the spectrum of states is linear in the radial and angular momentum quantum numbers:
$\mathcal{M} \sim 2 n \! + \! L$. In the soft-wall model a smooth infrared cutoff is chosen to model
confinement and reproduce the usual Regge behavior 
$\mathcal{M}^2 \sim n \! + \!  L$~\cite{Karch:2006pv}.
The resulting models, although {\it ad hoc}, provide a
simple semiclassical approximation to QCD which has both constituent counting
rule behavior at short distances and confinement at large distances~\cite{Brodsky:2008pg}.

It is thus natural, as a useful first approximation, to use the isometries of AdS to map the local interpolating operators at the UV boundary of AdS space to the modes propagating inside AdS.
The short-distance behavior of a hadronic state is
characterized by its twist  (dimension minus spin) 
$\tau = \Delta - \sigma$, where $\sigma$ is the sum over the constituent's spin
$\sigma = \sum_{i = 1}^n \sigma_i$. Twist is also equal to the number of partons $\tau = n$.
Under conformal transformations the interpolating operators transform according to their twist, and consequently the AdS isometries map the twist scaling dimensions into the AdS 
modes~\cite{Brodsky:2003px}. 

The eigenvalues of normalizable modes
in AdS give the hadronic spectrum. AdS modes represent also the probability
amplitude for the distribution of quarks and gluons at a given scale.
There are also non-normalizable modes which are related to 
external currents: they propagate into the AdS  interior  and couple to
boundary QCD interpolating operators~\cite{Gubser:1998bc, Witten:1998qj}.
Following this simplified ``bottom up" approach,  a limited set of operators is  introduced to construct 
phenomenological viable five-dimensional dual holographic 
models~\cite{Boschi-Filho:2002vd, deTeramond:2005su, Erlich:2005qh, DaRold:2005zs}. 

In the top-down supergravity approach, one introduces higher
dimensional branes to the ${\rm AdS}_5 \times  S^5$
background~\cite{Karch:2002sh} in order to have a theory 
of flavor. One can obtain models with massive quarks in the fundamental representation,  compute 
the hadronic spectrum, and describe chiral 
symmetry breaking  in the context of higher dimensional brane 
constructs~\cite{Karch:2002sh, Babington:2003vm, Kruczenski:2003uq,
Sakai:2004cn, Gursoy:2007er}.  However, a theory dual to QCD is unknown, and this ``top-down" approach is difficult to extend beyond theories exceedingly constrained by their
symmetries.

An important feature of light-front
quantization is the fact that it provides exact formulas for
current matrix elements as a sum of bilinear forms which can be mapped
into their AdS/CFT counterparts in the semiclassical approximation.
The AdS metric written in terms of light front coordinates $x^\pm =
x^0 \pm x^3$ is
\begin{equation} \label{eq:AdSzLF}
ds^2 = \frac{R^2}{z^2} \left( dx^+ dx^- - d \mbf{x}_\perp^2 - dz^2
\right).
\end{equation}
At fixed light-front time $x^+=0$, the metric depends only on the transverse
$ \mbf{x}_\perp$ and the holographic variable $z$.
Thus we can find an exact correspondence between the
fifth-dimensional coordinate of anti-de Sitter space $z$ and a
specific impact variable $\zeta$ in the light-front formalism.  The
new variable $\zeta$
measures the separation of the constituents within the hadron in
ordinary space-time.  The amplitude $\Phi(z)$ describing  the
hadronic state in $\rm{AdS}_5$ can then be precisely mapped to the
light-front wavefunctions $\psi_{n/H}$ of hadrons in physical
space-time~\cite{Brodsky:2006uqa, Brodsky:2007hb}, thus providing a relativistic
description of hadrons in QCD at the amplitude level. 

The correspondence of AdS amplitudes to the QCD wavefunctions in light-front coordinates
was carried out in~\cite{Brodsky:2006uqa, Brodsky:2007hb}
by comparing the expressions for the electromagnetic matrix elements in QCD and
AdS for any value of the momentum transfer $q^2$. It is indeed remarkable that 
such a correspondence exists, since strings describe extended objects
coupled to an electromagnetic field distributed in the AdS interior, whereas QCD
degrees of freedom are pointlike particles with individual local couplings to the electromagnetic
current. However, as we have shown~\cite{Brodsky:2006uqa, Brodsky:2007hb}, a precise mapping of AdS modes to hadronic light-front wavefunctions
can be found in the strongly-coupled semiclassical approximation to QCD.

The matrix elements of local operators of hadronic composite systems, such as  currents, angular momentum and the energy-momentum tensor,  have exact Lorentz invariant representations in the light front in terms of the overlap of light-front wave functions.  One may ask, if the holographic mapping found in~\cite{Brodsky:2006uqa, Brodsky:2007hb} for the electromagnetic current
is specific to the charge distribution within a hadron or a general feature of
light-front AdS/QCD. 

The matrix elements of the energy-momentum tensor $\Theta^{\mu \nu}$ of each constituent define the gravitational form factor of a composite hadron.
In this paper we shall use gravitational matrix elements to obtain the holographic mapping of the AdS mode  wavefunctions $\Phi(z)$ in  AdS space to the light-front wavefunctions 
$\psi_H$ in physical $3+1$ space-time  defined at fixed light-cone time $\tau= t + z/c$.   We find the identical holographic mapping from $z \to \zeta$ as in the electromagnetic case. The agreement of the results for electromagnetic and gravitational hadronic transition amplitudes provides an important consistency test and verification of holographic mapping from AdS to physical observables defined on the light-front. 

This paper is organized as follows. After briefly reviewing the QCD light-front Fock representation
in Sec.  \ref{LFFrep}, we derive in Sec. \ref{GFF-QCD} the exact form of matrix elements of the energy-momentum tensor
for a $n$-parton composite object in light-front QCD. In Sec. \ref{GFF-AdS} we discuss the 
gravitational form factors in AdS/QCD. In Sec. \ref{TME-AdS} we describe  the normalization of the AdS hadronic solutions to the energy-momentum tensor and obtain
the corresponding hadronic transition matrix
elements in AdS space. The actual mapping from AdS to QCD matrix elements 
is carried out in Sec. \ref{LFM},  where the Hamiltonian in the holographic  light-front
representation is related to the light-front Schr\"odinger equation predicted from AdS/QCD. 
Some final remarks are given in the conclusions in Sec. \ref{Conclusions}. 
Other aspects useful for the discussion of the paper are given in the appendices.
In particular we describe in Appendix \ref{PionLFM} the specific AdS/QCD mapping for a  two-parton hadronic bound state, which is useful for understanding the $n$-parton results discussed in this article.

\section{Light-Front Fock Representation}
\label{LFFrep}

The light-front expansion of any hadronic system
is constructed by quantizing QCD
at fixed light-cone time \cite{Dirac:1949cp} $\tau = t + z/c$.
In terms of the hadron  four-momentum $P = (P^+, P^-, \mbf{P}_{\!\perp})$,
$P^\pm = P^0 \pm P^3$,
the light-cone Lorentz invariant Hamiltonian for the composite system
$H_{LF}^{QCD} = P^-P^+ - \mbf{P}^2_\perp$  has
eigenvalues given in terms of the eigenmass ${\cal M}$ squared  corresponding 
to the mass spectrum of the color-singlet states in QCD~\cite{Brodsky:1997de}
\begin{equation}
H_{LF} \vert{\psi_H}\rangle = {\cal M}^2_H \vert{\psi_H}\rangle,
\end{equation}
where $\vert{\psi_H}\rangle$ is an expansion in multiparticle Fock eigenstates
$\{\vert{n} \rangle\}$ of the free light-front (LF) Hamiltonian: 
$\vert \psi_H \rangle = \sum_n \psi_{n/H} \vert \psi_H\rangle $.
The light-front wave functions (LFWFs) $\psi_{n/H}$ provide a
{\it frame-independent } representation of a hadron which relates its quark
and gluon degrees of freedom to their asymptotic hadronic state.

The hadron wavefunction is an eigenstate of the total momentum $P^+$
and $\mbf{P}_{\! \perp}$ and the longitudinal spin projection $S_z$,
and is normalized according to
\begin{multline}
\bigl\langle \psi_H(P^+,\mbf{P}_{\! \perp}, S_z) \big\vert 
\psi_H(P'^+,\mbf{P}'_\perp,S_z') \bigr\rangle \\ 
= 2 P^+ (2 \pi)^3 \,\delta_{S_z,S'_z} \,\delta \bigl(P^+ - P'^+ \bigr)
\,\delta^{(2)} \negthinspace \bigl(\mbf{P}_{\! \perp} - \mbf{P}'_\perp\bigr) . 
\label{eq:Pnorm}
\end{multline}

Each hadronic eigenstate $\vert \psi_H \rangle$  is expanded in
a Fock-state complete basis of noninteracting $n$-particle states
$\vert n \rangle$ with an infinite number of components
\begin{multline}
\left\vert \psi_H(P^+,\mbf{P}_{\! \perp}, S_z) \right\rangle  \\
= \sum_{n,\lambda_i} 
\prod_{i=1}^n \int \! \frac{dx_i}{\sqrt{x_i}}
\frac{d^2 \mbf{k}_{\perp i}}{2 (2\pi)^3} \, 16 \pi^3 \,
\delta \Bigl(1 - \sum_{j=1}^n x_j\Bigr) 
\delta^{(2)} \negthinspace\Bigl(\sum_{j=1}^n\mbf{k}_{\perp j}\Bigr) \\
\times \psi_{n/H}(x_i,\mbf{k}_{\perp i},\lambda_i) 
\bigl\vert n: x_i P^+\!, x_i \mbf{P}_{\! \perp} \! + \! \mbf{k}_{\perp i},\lambda_i \bigr\rangle,
\label{eq:LFWFexp}
\end{multline}
where the sum begins with the valence state; e.g., $n \ge 3$ for baryons. The
coefficients of the  Fock expansion
\begin{equation} \label{eq:LFWF}
\psi_{n/H}(x_i, \mbf{k}_{\perp i},\lambda_i) 
= \bigl\langle n:x_i,\mbf{k}_{\perp i},\lambda_i \big\vert \psi_H\bigr\rangle ,
\end{equation}
are independent of the total momentum $P^+$ and $\mbf{P}_{\! \perp}$ of
the hadron and depend only on the relative partonic coordinates,
the longitudinal momentum fraction $x_i = k_i^+/P^+$,
the relative transverse momentum $\mbf{k}_{\perp i}$
and $\lambda_i$, the
projection of the constituent's spin along the $z$ direction. 
Thus, given the Fock projection (\ref{eq:LFWF}), the wavefunction
of a hadron is determined in any frame. The
amplitudes $\psi_{n/H}$ represent the probability amplitudes to find
on-mass-shell constituents $i$ with longitudinal momentum $ x_i P^+$, 
transverse momentum $x_i \mbf{P}_{\! \perp} + \mbf{k}_{\perp i}$ and
helicity $\lambda_i$ in the hadron $H$. Momentum conservation requires
$\sum_{i=1}^n x_i =1$ and $\sum_{i=1}^n \mbf{k}_{\perp i}=0$.
In addition, each light front wavefunction
$\psi_{n/H}(x_i,\mbf{k}_{\perp i},\lambda_i)$ obeys the angular momentum sum 
rule~\cite{Brodsky:2000ii}
$J^z  = \sum_{i=1}^n  S^z_i + \sum_{i=1}^{n-1} L^z_i $,
where $S^z_i = \lambda_i $ and the $n-1$ orbital angular momenta
have the operator form 
\begin{equation} \label{eq:L}
L^z_i =-i \left(\frac{\partial}{\partial k^x_i}k^y_i -
\frac{\partial}{\partial k^y_i}k^x_i \right).
\end{equation}
It should be emphasized that the assignment of quark and gluon spin and orbital angular momentum of a hadron is  a gauge-dependent concept. The LF framework in light-cone gauge $A^+=0$ provides a physical definition since there are no gauge field ghosts and the gluon 
has spin-projection $J^z= \pm 1$; moreover, it is frame-independent.

The LFWFs are normalized according to
\begin{equation}
\sum_n  \int \big[d x_i\big] \left[d^2 \mbf{k}_{\perp i}\right]
\,\left\vert \psi_{n/H}(x_i, \mbf{k}_{\perp i}) \right\vert^2 = 1,
\label{eq:LFWFnorm}
\end{equation}
where the measure of the constituents phase-space momentum
integration  is
\begin{equation}
\int \big[d x_i\big] \equiv
\prod_{i=1}^n \int dx_i \,\delta \Bigl(1 - \sum_{j=1}^n x_j\Bigr) ,
\end{equation}
\vspace{-10pt}
\begin{equation}
\int \left[d^2 \mbf{k}_{\perp i}\right] \equiv \prod_{i=1}^n \int
\frac{d^2 \mbf{k}_{\perp i}}{2 (2\pi)^3} \, 16 \pi^3 \,
\delta^{(2)} \negthinspace\Bigl(\sum_{j=1}^n\mbf{k}_{\perp j}\Bigr).
\end{equation}
The spin indices have been suppressed.

The complete basis of Fock-states $\vert{n}\rangle$ is constructed by applying 
free-field creation operators to
the vacuum state $\vert 0 \rangle$ which has no particle content,
$P^+ \vert 0 \rangle =0$, $\mbf{P}_{\! \perp} \vert 0 \rangle = 0$.  The
fundamental constituents appear in light-front quantization as the
excitations of the dynamical fields, the Dirac field $\psi_+$,
$\psi_\pm = \Lambda_\pm \psi$, $\Lambda_\pm = \gamma^0 \gamma^\pm$,
and the transverse field $\mbf{A}_\perp$ in the $A^+ = 0$ gauge, each
expanded in terms of quark and gluon creation and annihilation operators on the
transverse plane with coordinates $x^- = x^0 - x^3$ and $\mbf{x}_\perp$ 
at fixed light-front time $x^+ = x^0 +
x^3$~\cite{Brodsky:1997de}.  For each kind of quark $f$ the Dirac field operator is expanded as
\begin{multline} \label{eq:psiop}
\psi_f(x)_\alpha = \sum_\lambda \int_{q^+ > 0} \frac{d q^+}{\sqrt{ 2
 q^+}}
\frac{d^2 \mbf{q}_\perp}{ (2 \pi)^3} \\ \times
\left[b^f_\lambda (q)
u_\alpha(q,\lambda) e^{-i q \cdot x} + d^f_\lambda (q)^\dagger
v_\alpha(q,\lambda) e^{i q \cdot x}\right],
\end{multline}
with commutation relations
\begin{multline} \label{eq:cr}
\left\{b(q), b^\dagger(q')\right\} = \left\{d(q), d^\dagger(q')\right\}  \\
= (2 \pi)^3 \,\delta (q^+ - {q'}^+)
\delta^{(2)}\negthinspace\left(\mbf{q}_\perp - \mbf{q}'_\perp\right) .
\end{multline}
Similar expansion follow for the
transverse gluon field $\mbf{A}_\perp$.
We shall use the Lepage-Brodsky (LB) conventions~\cite{Lepage:1980fj} for the
properties of the light-cone spinors.
A one-particle state is defined by
$\vert q \rangle = \sqrt{2 q^+} \,b^\dagger(q) \vert 0 \rangle$.
Each n-particle Fock state
$|p_i^+, \mbf{p}_{\perp i} \rangle$ is an eigenstate of  $P^+$ and
$\mbf{P}_{\! \perp}$ and is normalized according to
\begin{multline}
\bigl\langle  p_i^+, \mbf{p}_{\perp i},\lambda_i
\big|{p'}_i^+, \mbf{p'}_{\negthinspace\perp i},\lambda_i' \bigr\rangle \\
=  2 p_i^+ (2 \pi)^3 \, \delta \bigl(p_i^+ - {p'}_i^+\bigr) \, 
\delta^{(2)} \negthinspace \bigl(\mbf{p}_{\perp i} - \mbf{p'}_{\negthinspace\perp
    i}\bigr) \, \delta_{\lambda_i,\lambda_i'}.
\label{eq:normFC}
\end{multline}

The LFWFs $\psi_n(x_j, \mbf{k}_{\perp j})$ can be expanded in terms of  $n-1$ independent
transverse coordinates $\mbf{b}_{\perp j}$,  $j = 1,2,\dots,n-1$,
conjugate to the relative coordinates $\mbf{k}_{\perp i}$
\begin{multline} \label{eq:LFWFb}
\psi_n(x_j, \mathbf{k}_{\perp j}) =  (4 \pi)^{(n-1)/2} 
\prod_{j=1}^{n-1}\int d^2 \mbf{b}_{\perp j} \\ \times
\exp{\Big(i \sum_{k=1}^{n-1} \mathbf{b}_{\perp k} \cdot \mbf{k}_{\perp k}\Big)} \,
\tilde{\psi}_n(x_j, \mathbf{b}_{\perp j}),
\end{multline}
where $\sum_i \mbf{b}_{\perp i} = 0$. The normalization is defined by
\begin{equation}  
\sum_n  \prod_{j=1}^{n-1} \int d x_j d^2 \mbf{b}_{\perp j} 
\left\vert\tilde \psi_n(x_j, \mbf{b}_{\perp j})\right\vert^2 = 1.
\end{equation}

\section{Gravitational Form Factors of Composite Hadrons in QCD}
\label{GFF-QCD}

Matrix elements of the energy-momentum tensor $\Theta^{\mu \nu} $ which define the gravitational form factors play an important role in hadron physics.  Since one can define $\Theta^{\mu \nu}$ for each parton, can one can identify the momentum fraction and  contribution to the orbital angular momentum of each quark flavor and gluon of a hadron. For example, the spin-flip form factor $B(q^2)$, which is the analog of the Pauli form factor $F_2(Q^2)$ of a nucleon, provides a  measure of the orbital angular momentum carried by each quark and gluon constituent of a hadron at $q^2=0.$   Similarly,  the spin-conserving form factor $A(q^2)$, the analog of the Dirac form factor $F_1(q^2)$, allows one to measure the momentum  fractions carried by each constituent.
This is the underlying physics of Ji's sum rule~\cite{Ji:1996ek}:
$\langle J^z\rangle = \half [ A(0) + B(0)] $,  which has prompted much of the current interest in 
the generalized parton distributions (GPDs)  measured in deeply
virtual Compton scattering~\cite{Kumericki:2008di}. Measurements of the GPD's are of particular relevance
for determining the distribution of partons in the transverse
impact plane, and thus could be confronted with AdS/QCD predictions which follow
from the mapping of AdS modes to the transverse impact representation~\cite{Brodsky:2006uqa}.

An important constraint is $B(0) = \sum_i B_i(0) = 0$;  i.e.  the anomalous gravitomagnetic moment of a hadron vanishes when summed over all the constituents $i$. This was originally derived from the equivalence principle of gravity~\cite{Teryaev:1999su}. The explicit verification of these relations, Fock state by Fock state, can be obtained in the light-front quantization of QCD in  light-cone gauge~\cite{Brodsky:2000ii}.  Physically $B(0) =0$ corresponds to the fact that the sum of the $n$ orbital angular momenta $L$ in an $n$-parton Fock state must vanish since there are only $n-1$ independent orbital angular momenta (\ref{eq:L}).

Gravitational form factors can also be computed in AdS/QCD from the overlap integral of
hadronic string modes propagating in AdS space with a graviton field $h_{\mu \nu}$ 
which acts as a source and probes the AdS interior. This has  been done very recently for the gravitational form factors of mesons  by Abidin and Carlson~\cite{Abidin:2008ku}, thus providing restrictions on the GPDs. 

Recent applications to the electromagnetic form factors  of hadrons~\cite{Brodsky:2008pg, Brodsky:2007hb, Grigoryan:2007vg, Kwee:2007dd, Agaev:2008zz}   
in the bottom-up and in the top-down 
string framework~\cite{RodriguezGomez:2008zp}
of the AdS/CFT correspondence
have followed from the original papers~\cite{Polchinski:2002jw,Hong:2004sa}. 
Here we shall extend our previous results~\cite{Brodsky:2006uqa, Brodsky:2007hb} for the holographic mapping of AdS current matrix elements to gravitational form factors. 
If both quantities for the gravitational form factors represent the same physical observable for
any value of the momentum transfer $q^2$, then an exact correspondence can be established
between the AdS modes $\Phi(z)$ and LFWFs of hadrons $\psi_{n/H}$ as in the case
of the electromagnetic form factors. To simplify the discussion, we will consider the holographic mapping of matrix
elements of the energy-momentum tensor of mesons, where only one gravitational form factor is present,
but the results can be extended to other hadrons as shown in~\cite{Abidin:2008ku}.

The QCD Lagrangian density is
\begin{equation}
\mathcal{L}_{\rm QCD} = \bar \psi \left( i \gamma^\mu D_\mu - m\right) \psi 
- \tfrac{1}{4} G^a_{\mu \nu} G^{a \, \mu \nu} ,
\end{equation}
where $D_\mu = \partial_\mu - i g_s A^a_\mu T^a$ and 
$G^a_{\mu \nu} = \partial_\mu A^a_\nu - \partial_\nu A^a_\mu + 
g_s c^{abc} A_\mu^b A_\nu^c$, with $\left[T^a, T^b\right] = i c^{abc} T^c$ and
$a, b ,c$ are $SU(3)$ color indices.

We can find a symmetric and gauge-invariant
expression for the energy-momentum tensor $\Theta_{\mu \nu}$,
the Hilbert  energy-momentum tensor, by varying  the QCD action with respect to the four-dimensional Minkowski space-time metric ${\rm g}_{\mu \nu}(x)$ 
\begin{equation}
\Theta_{\mu \nu}(x) = - \frac{2}{\sqrt{\rm g}} \frac{\delta S_{\rm QCD}}{\delta {\rm g}^{\mu \nu}(x) }  ,
\end{equation}
where $S_{\rm QCD} = \int \! d^4 x \sqrt{\rm g} \, \mathcal{L}_{\rm QCD}$
and ${\rm g} \equiv \vert {\rm det} \, {\rm g}_{\mu \nu} \vert$.
The result is
\begin{multline} \label{eq:emt}
\Theta_{\mu \nu} =  \tfrac{1}{2}  
\bar \psi  i \! \left( \gamma_\mu D_\nu + \gamma_\nu D_\mu \right) \psi
- {\rm g}_{\mu \nu} \bar \psi \left( i \Dslash - m\right)\psi  \\
 - G^a_{\mu \lambda} {G^a_\nu}^{\hspace{0.5pt} \lambda}
 + \tfrac{1}{4} {\rm g}_{\mu\nu}  G^a_{\lambda \sigma} G^{a \hspace{1pt} \lambda \sigma}.
\end{multline}
The first two terms in (\ref{eq:emt}) correspond to the fermionic contribution to the
energy-momentum tensor  and the last two to the gluonic contribution.  In terms of
(\ref{eq:emt}) the total angular momentum operator $J$ of the composite hadron can
be expressed in the gauge-invariant form
\begin{equation}
J_i = \half \epsilon_{ijk} \!  \int d^3x \!
\left[ \Theta^{0k} x^j - \Theta^{0j} x^k \right].
\end{equation}

In the semiclassical AdS/CFT correspondence there are no quantum effects, and 
only the valence Fock state contributes to the hadronic wave function. In this
approximation we need to consider only the quark contribution to the energy 
momentum tensor. In the light-front
gauge $A^+ = 0$ the fermionic component $\Theta^{++}$ is
\begin{equation} \label{eq:LFemt}
\Theta^{++}(x) =  \frac{i}{2} \sum_f \bar \psi_f(x) \gamma^+ 
\overleftrightarrow\partial^{\!+} \psi_f(x),
\end{equation}
where an integration by parts is carried out to write $\Theta^{++}$ in its hermitian operator 
form. The sum 
in (\ref{eq:LFemt}) extends over all the types of quarks $f$ present in the hadron.
Notice that the second term of the energy-momentum tensor (\ref{eq:emt}) does
not appear in the expression for $\Theta^{++}$ since the metric component ${\rm g}^{++}$ 
is zero in the light-front as discussed in Appendix \ref{Metric}.

We will use light-front frame coordinates
\begin{eqnarray} \label{eq:qframe}
P  &=& (P^+, P^-, \mbf{P}_{\! \perp}) = \Bigl( P^+,\frac{ M^2}{ P^+},
\vec 0_{\perp} \Bigr),\\q  &=& (q^+, q^-, \mbf{q}_{\perp}) = \Bigl(
0,  \frac{2 \,q \! \cdot \! P}{P^+}, \mbf{q}_{\perp} \Bigr), \nonumber
\end{eqnarray}
where  $q^2 = - Q^2 = -2 \,q \! \cdot \! P
 = - \mbf{q}^2_\perp$ is the spacelike
four-momentum squared transferred to the composite hadron. The
gravitational form factor of a meson is defined in terms of 
matrix elements of the ``plus-plus" components
 of the energy-momentum tensor evaluated at
light-cone time $x^+ = 0$. In the $q^+=0$ frame
\begin{equation}
\left\langle P' \left\vert \Theta^{++}(0) \right\vert P\right\rangle = 2
(P^+)^2  A(Q^2),
\end{equation}
where $P' = P + q$ and the gravitational form factor $A(Q^2)$ satisfy the momentum sum 
rule $A(0)  = 1$.

The expression for the operator $\Theta^{++}(0)$ in the particle number representation
follows from the momentum expansion of the Dirac field $\psi(x)$ in terms of creation
and annihilation operators given by (\ref{eq:psiop}). Using the light-cone metric conventions given
in Appendix \ref{Metric} and the results listed in Appendix A of \cite{Brodsky:2007hb} for the quark spinor transitions, we find
\begin{multline} \label{eq:EMTO}
\Theta^{++} =  \half \sum_{f, \lambda}
\int \frac{d q^+ d^2\mbf{q}_\perp}{(2 \pi)^3} 
\int \frac{d q'^+ d^2\mbf{q}'_\perp}{(2 \pi)^3}  
\left(q^+ \! + q'^+ \right) \\ \times
\bigl\{ b_\lambda^{f \dagger}(q) b^f_\lambda(q')
     + d_\lambda^{f \dagger}(q')  d^f_\lambda(q) \bigr\}.
\end{multline}
The operator $\Theta^{++}$ annihilates a quark (antiquark) 
with momentum $q'$ ($q$) and spin projection $\lambda$ along the $z$
direction and creates a quark (antiquark) with the same spin and
momentum $q$ ($q'$).

The matrix element of the energy momentum tensor
$\left\langle\psi_{P'}  \left \vert \Theta^{++}(0) \right\vert \psi_P \right\rangle$ can be computed
by expanding the initial and final hadronic states
in terms of its Fock components using (\ref{eq:LFWFexp}). The transition
amplitude can then be expressed as a sum of overlap integrals with diagonal $\Theta^{++}$-matrix
elements in the $n$-particle Fock-state basis.
For each Fock state, we label with $i = n$ the struck
constituent quark with light-front longitudinal momentum fraction $x_n = x$ and 
with $j = 1, 2, \dots, n-1$ each
spectator with longitudinal momentum fraction $x_j$. Using the normalization condition
(\ref{eq:normFC}) for each individual constituent and after
integration over the intermediate variables in the $q^+ = 0$ frame,
we find the expression  for the gravitational form factor of a meson~\cite{Brodsky:2000ii}
\begin{multline} \label{eq:MGFF}
A(q^2) = \sum_n \int \big[d x_i\big] \left[d^2 \mbf{k}_{\perp i}\right]   \\ \times
\sum_{f=1}^n x_f  \,
 \psi^*_{n/H} (x_i, \mbf{k}'_{\perp i},\lambda_i)
\psi_{n/H} (x_i, \mbf{k}_{\perp i},\lambda_i),
\end{multline}
where the sum is over all the partons in each Fock state $n$. 
The variables of the light-cone Fock components in the
final-state are given by $\mbf{k}'_{\perp i} = \mbf{k}_{\perp i} 
+ (1 - x_i)\, \mbf{q}_\perp $ for a struck  constituent quark and 
$\mbf{k}'_{\perp i} = \mbf{k}_{\perp i} - x_i \, \mbf{q}_\perp$ for each
spectator. 
Notice that each type of parton
contributes to the gravitational form factor with struck constituent light-cone momentum fractions $x_f$, 
instead of the electromagnetic constituent charge $e_f$ which appears in the electromagnetic form factor.
Since the longitudinal momentum fractions of the constituents add to one, $\sum_f x_f = 1$,
the momentum sum rule is satisfied at $q =0$: $A(0) = 1$;
the formulae are exact if the sum is over all Fock states $n$. 
Notice that there is a factor of $N_C$ from a closed quark loop where the graviton
is attached and a normalization factor of $1/\sqrt{N_C}$ for each
meson wave function; thus color factors cancel out from the
expression of the gravitational form factor.

In the light-front formalism matrix elements of local operators
are  represented  as overlaps of light-front wavefunctions. 
In order to compare with AdS results it is  convenient to express the LF
expressions in the transverse impact representation since the
bilinear forms may be expressed in terms of the product of light-front wave
functions  with identical variables. 
We substitute (\ref{eq:LFWFb}) in the formula (\ref{eq:MGFF}). 
Integration over $k_\perp$ phase space gives us $n - 1$ delta
functions to integrate over the $n - 1$ intermediate transverse variables
with the result
\begin{multline} \label{eq:MGFFbb} 
A(q^2) =  \sum_n \prod_{j=1}^{n-1}\int d x_j d^2 \mbf{b}_{\perp j}  \\  \times \sum_{f=1}^n x_f
\exp \! {\Bigl(i \mbf{q}_\perp \! \cdot \sum_{k=1}^{n-1} x_k \mbf{b}_{\perp k}\Bigr)} 
\left\vert \tilde \psi_n(x_j, \mbf{b}_{\perp j}, \lambda_j)\right\vert^2,
\end{multline}
corresponding to a change of transverse momentum $x_j \mbf{q}_\perp$ for each
of the $n-1$ spectators and is valid for any Fock state $n$. The results can be summed over 
$n$ to obtain an exact representation.

\subsection{Effective Single-Particle Distribution}
\label{ESPD}

We can define $A_{f/n}(q^2)$ which is the contribution to the gravitational form-factor from the struck parton $f$ in Fock state $n$. In terms of the $n-1$ independent coordinates $x_k$ and 
$\mbf{b}_{\perp k}$, $k = 1, 2, \dots, n, ~ k \ne f$ we have
\begin{multline}
A_{f/n}(q^2) =  \prod_{k \ne f} \int  \! d x_k \, d^2 \mbf{b}_{\perp k}   
\, \Big(1 - \sum_{\ell \ne f} x_\ell  \Big) \\ \times
\exp \! {\Big(i \mbf{q}_\perp \! \cdot \sum_{m \ne f} x_m\mbf{b}_{\perp m}\Big)} 
\left\vert \tilde \psi_n \big(x_k, \mbf{b}_{\perp k}\big)\right\vert_{k \ne f}^2.
\end{multline}

Following~\cite{Brodsky:2006uqa, Brodsky:2007hb}  we can write the gravitational form factor 
in terms of an effective single particle density~\cite{Soper:1976jc} in the light-front frame.
Summing over Fock states $A_f(q^2) = \sum_n A_{f/n}(q^2)$, we have
\begin{equation}
A_f(q^2) = \! \int_0^1 \!   x\,  dx \, \rho_f(x, \mbf{q}_\perp),
\end{equation}
where $A(0) = \sum_f A_f(0) = \sum_f \langle x_f \rangle = 1$.
The effective density $\rho_f(x, \mbf{q}_\perp)$ is given by
\begin{multline} \label{eq:rhoxq}
\rho_f(x, \mbf{q}_\perp) = \sum_n \prod_{k \ne f}
\int dx_k \, d^2 \mbf{b}_{\perp k}  \, \delta \Bigl(1-x - \sum_{\ell \ne f} x_\ell\Bigr) \\ \times
 \exp{ \!\Bigl(i \mbf{q}_\perp \!
\cdot \sum_{m \ne f} x_m \mbf{b}_{\perp m}\Bigr)}
\left\vert\tilde \psi_n(x_k, \mbf{b}_{\perp k})\right\vert_{k \ne f}^2.
\end{multline}
The integration is over the coordinates of the $n-1$ spectator
partons, and  $x $ is the coordinate of the active 
quark with longitudinal momentum $x$.
We can also write the form factor in terms of an effective  single-particle 
transverse distribution $\tilde\rho_f(x, \vec \eta_\perp)$
\begin{equation} \label{eq:FFfeta}
A_f(q^2) = 
 \! \int^1_0 \!  x  \, dx  \! \int  \! d^2 \vec \eta_\perp e^{i \vec \eta_\perp \cdot \mbf{q}_\perp}
\tilde\rho_f(x,\vec \eta_\perp),
\end{equation}
where
$\vec \eta_\perp = \sum_{k \ne f} x_k \mbf{b}_{\perp k}$
is the $x$-weighted transverse position coordinate of the $n-1$
spectators. The corresponding transverse density is~\cite{Brodsky:2006uqa, Brodsky:2007hb}
\begin{multline} \label{eq:rhoeta}
\tilde\rho_f(x, \vec \eta_\perp) = 
\int \! \frac{d^2 \mbf{q}_\perp}{(2 \pi)^2}  \,
e^{-i \vec \eta_\perp \cdot \mbf{q}_\perp} \rho_f(x, \mbf{q}_\perp) \\ 
= \sum_n \prod_{k \ne f} \int \! dx_k \, d^2\mbf{b}_{\perp k} 
\,\delta \Bigl(1-x -\sum_{\ell \ne k} x_\ell\Bigr) \\ \times
\delta^{(2)}\Bigl(\sum_{m \ne f} x_m \mbf{b}_{\perp m} - \vec \eta_\perp\Bigr)
\left\vert \tilde\psi_n(x_k, \mbf{b}_{\perp k}) \right\vert_{k \ne f}^2.
\end{multline}
It is useful to integrate (\ref{eq:FFfeta}) over angle; we obtain
\begin{multline} \label{eq:GFFzeta} 
A(q^2) = 2 \pi \! \sum_f \int_0^1\!   dx \, (1-x) \\ \times \int \!  \zeta d \zeta 
J_0 \! \left(\! \zeta q \sqrt{\frac{1-x}{x}}\right)  \! \tilde\rho_f(x,\zeta),
\end{multline}
where we have introduced the variable
\begin{equation} \label{eq:zeta}
\zeta = \sqrt{\frac{x}{1-x}} \, \Big\vert \sum_{k \ne f} x_k \mathbf{b}_{\perp k}\Big\vert,
\end{equation}
representing the $x$-weighted transverse impact coordinate of the
spectator system.

\section{Gravitational Form Factors in AdS/CFT}
\label{GFF-AdS}

AdS coordinates are the $d=4$ Minkowski coordinates
$x^\mu$ and $z$, the holographic coordinate, which we label 
$x^\ell = (x^\mu, z)$. The metric of  AdS$_{d+1}$
space-time is
\begin{equation} \label{eq:gAdS}
ds^2 = \frac{R^2}{z^2} \left(\eta_{\mu \nu} dx^\mu dx^\nu - dz^2\right),
 \end{equation}
where the AdS radius is $R$. 
Fields propagating in 5-dimensional AdS space
are represented by capital letters such as
$\Phi$. Holographic modes in 4-dimensional Minkowski space are represented by
$\phi$. 

\subsection{Gauge/Gravity Semiclassical Correspondence}

The formal statement of the duality between a gravity theory on $(d+1)$-dimensional 
Anti-de Sitter AdS$_{d+1}$ space and the strong coupling limit of a conformal
field theory (CFT)  on the $d$-dimensional asymptotic boundary of AdS$_{d+1}$
at $z=0$ is expressed in terms of the $d+1$ partition function for
a field $\Phi(x,z)$ propagating in the bulk
\begin{equation}
Z_{grav}[\Phi(x,z)] = e^{i S_{eff}[\Phi]} =
\int \mathcal{D}[\Phi] e^{ i S[\Phi]},
\label{eq:Zgrav}
\end{equation}
where $S_{eff}$ is the effective action of the AdS$_{d+1}$ theory,
and the $d$-dimensional generating functional 
of the conformal field theory in presence of an external source $\Phi_0(x)$,
\begin{multline} 
  Z_{CFT}[\Phi_0(x)] =  e^{ i W_{CFT}[\Phi_0]} \\
  = \left< \exp\left(i \int d^dx  \Phi_0(x) \mathcal{O}(x)\right) \right>.
\label{eq:ZCFT}
\end{multline}
The functional $W_{CFT}$ is the generator of connected 
Green's functions of the boundary theory and $\mathcal{O}(x)$ is a QCD local
interpolating operator.
The precise relation of the gravity theory on AdS space
to the conformal field theory at its boundary 
is~\cite{Gubser:1998bc, Witten:1998qj}
\begin{equation}
 Z_{grav}\big[\Phi(x,z) \vert_{z = 0} = \Phi_0(x) \big]
 = Z_{CFT}\left[\Phi_0\right],
\label{eq:grav-CFT}
\end{equation}
where the partition function (\ref{eq:Zgrav}) on AdS$_{d+1}$ is integrated 
over all possible configurations
$\Phi$ in the bulk which approach its boundary value $\Phi_0$.
If we neglect the contributions from the nonclassical configurations 
to the gravity partition function, then the functional $W_{CFT}$  of the four-dimensional gauge theory
(\ref{eq:ZCFT}) is precisely equal to the classical (on-shell) gravity action
(\ref{eq:Zgrav})
\begin{equation}
W_{CFT}\left[\phi_0\right] =
S_{eff}\big[\Phi(x,z)\vert_{z=0} = \Phi_0(x)\big]_{\rm on-shell},
\end{equation}
evaluated in terms of the classical solution to the bulk equation of motion.
This defines  the semiclassical approximation to the conformal field theory.
In the limit $z \to 0$, the independent solutions behave as
\begin{equation} \label{eq:Phiz0}
\Phi(x,z) \to z^\Delta \,\Phi_+(x) + z^{d - \Delta} \,\Phi_-(x),
\end{equation}
where $\Delta$ is the conformal dimension.
The non-normalizable solution $\Phi_-$  is the boundary value of the bulk
field $\Phi$ which couples to a QCD gauge-invariant operator 
$\mathcal{O}$ in the $z \to 0$ asymptotic boundary, thus $\Phi_- = \Phi_0$.
The normalizable solution $\Phi_+(x)$ is the response
function and corresponds to the physical states~\cite{Balasubramanian:1998sn}. 
The  interpolating
operators $\mathcal{O}$ of the boundary conformal theory are constructed
from local gauge-invariant products of quark and gluon fields and
their covariant derivatives, taken at the same point in
four-dimensional space-time in the $x^2 \to 0$ limit. 
Their conformal twist-dimensions  are matched to the  scaling behavior of
the AdS fields in the limit $z \to 0$ and are thus encoded into
the propagation of the modes inside AdS space. 

Integrating by parts and using the equation of motion for a scalar field in AdS
(as discussed below), the bulk
contribution to the action vanishes, and one is left with a non-vanishing surface
term in the ultraviolet boundary
\begin{equation} 
S = R^{d-1} \lim_{z \to 0} 
\int d^dx \, \frac{1}{z^{d-1}} \, \Phi \partial_z \Phi, 
\label{eq:SUV}
\end{equation}
which can be identified with the boundary functional $W_{CFT}$. 
Substituting the leading dependence (\ref{eq:Phiz0}) of $\Phi$ near $z =0$  in the ultraviolet
surface action (\ref{eq:SUV}) and using the functional relation
\begin{equation}
\frac{\delta W_{CFT} }{\delta \Phi_0} = \frac{\delta S_{\rm eff}}{\delta\Phi_0},
\end{equation}
it follows that
$\Phi_+(x)$ is related to the expectation values of $\cal O$
in the presence of the source $\Phi_0$~\cite{Balasubramanian:1998sn}
\begin{equation} \label{eq:DimPhi}
\left\langle 0 \vert {\cal O}(x) \vert 0 \right\rangle_{\Phi_0}
\sim \Phi_+(x).
\end{equation}
The exact relation depends on the normalization of the fields 
used~\cite{Klebanov:1999tb}. The field $\Phi_+$ thus acts as a
classical field, and it is the boundary limit of the normalizable string
solution which propagates in the bulk.

\subsection{Gravity Action}

The action for gravity coupled to a scalar field in AdS$_{d+1}$ space is
\begin{equation} 
S =  \int \! d^{d+1} x  \sqrt{g} \,
\Big(  \frac{1}{\kappa^2} \left(\mathcal{R} - 2 \Lambda\right) 
+ g^{\ell m} \partial_\ell \Phi^*\partial_m \Phi 
-  \mu^2 \Phi^* \Phi \Big),
\label{eq:SAdS}
\end{equation}
where $\mathcal{R}$ is the scalar curvature, $\kappa$ is the $d+1$ dimensional Newton constant 
and $\mu$ is a $d+1$ dimensional mass. The action is written as a sum of two terms
$S = S_G + S_M$,  where $S_G$
\begin{equation}
S_G =   \frac{1}{\kappa^2} \int \! d^{d+1} x  \sqrt{g} \,
  \left(\mathcal{R} - 2 \Lambda\right) ,
\label{eq:SAdSG}
\end{equation}
describes the  dynamics of the gravitational fields $g_{\ell m}$ and determines the AdS  background. The dynamics of all other fields, the matter fields, is included in $S_M$.  In the present discussion the matter content is represented by $\Phi$ and the action
\begin{equation}
S_M =  \int \! d^{d+1} x  \sqrt{g} \,
\left( g^{\ell m} \partial_\ell \Phi^*\partial_m \Phi 
-  \mu^2 \Phi^* \Phi \right),
\label{eq:SAdSM}
\end{equation}
describes a pion mode which propagates in AdS space.

The variation of the action with respect to the metric tensor gives Einstein's equations in the
presence of a bulk cosmological constant $\Lambda$:
\begin{equation}  \label{eq:EGE}
\mathcal{R}_{\ell m} - \half g_{\ell m} \mathcal{R} - \Lambda g_{\ell m} = 0.
\end{equation}
AdS space is a maximally symmetric space with Riemann tensor $R_{i k \ell m}$ 
\begin{equation}
\mathcal{R}_{i k \ell m} = - \frac{1}{R^2}
\left(g_{i \ell} g_{k m} - g_{i m} g_{k \ell}\right).
\end{equation}
By contracting $\mathcal{R}_{i k \ell m}$ we obtain the Ricci tensor 
$R_{i k} = g^{\ell m} R_{\ell i m k}$, 
\begin{equation}
R_{i k} = - \frac{d}{R^2} \, g_{i k}.
\end{equation}
Thus AdS space is an Einstein manifold. By further contracting the Ricci tensor
$\mathcal{R} = g^{i k} R_{i k} = g^{i \ell} g^{k m} \mathcal{R}_{i k \ell m}$,
we obtain the scalar curvature of AdS$_{d+1}$ space
$\mathcal{R} = - \frac{d(d+1)}{R^2}$,
a constant negative curvature.  From the equation of motion (\ref{eq:EGE}) we find the relation between
the cosmological constant and the AdS$_{d+1}$ radius
\begin{equation}
\Lambda = - \frac{d(d-1)}{2 R^2},
\end{equation}
thus $\Lambda = - \frac{6}{R^2}$ for $d= 4$.

Taking the variation of (\ref{eq:SAdSM}) with respect to $\Phi$ we find the AdS wave equation
for the pion mode
\begin{equation} \label{eq:AdSPhi}
z^3 \partial_z \left( \frac{1}{z^3} \partial_z \Phi \right)
- \partial_\rho  \partial^\rho \Phi - \left(\frac{\mu R}{z}\right)^2 \! \Phi= 0.
\end{equation}

\subsection{Interaction Terms in the Gravity Action}

The expression for the AdS matrix elements describing the interaction of the matter fields in AdS space with an external arbitrary  source at the AdS asymptotic boundary follows from the gauge-invariant definition of the energy-momentum tensor
\begin{equation} \label{eq:EMTMAdS}
\Theta_{\ell m}(x^\ell) = - \frac{2}{\sqrt{g}} \frac{\delta S_M}{\delta g^{\ell m}(x^\ell) }  ,
\end{equation}
where $g \equiv \vert {\rm det} \, g_{\ell m} \vert$.  In order to determine the precise form of
the transition amplitudes, we shall consider a small deformation of the metric  about its
AdS background $g_{\ell m}$ : $\bar g_{\ell m} = g_{\ell m} + h_{\ell m}$; 
we then expand $S_M$ to first order in $h_{\ell m}$. From 
(\ref{eq:SAdSM}) and (\ref{eq:EMTMAdS})
\begin{equation}
S_M[h_{\ell m}] = S_M[0] + \half \int \! d^{d+1}x \, \sqrt{g} \,  h_{\ell m} \Theta^{\ell m} + \mathcal{O}(h^2),
\end{equation}
where we have used the relation $\Theta^{\ell m} \delta g_{\ell m} = - \Theta_{\ell m} \delta g^{\ell m}$
which follows from $g^{\ell m} \delta g_{\ell m} = - g^{\ell m} \delta g_{\ell m}$.
Thus, in the weak gravitational approximation the coupling of an external graviton field $h_{\ell m}$
to matter is given by the interaction term ($d=4$)
\begin{equation} \label{eq:SI}
S_I =  \half \int \! d^4x \, dz \, \sqrt{g} \, h_{\ell m} \Theta^{\ell m},
\end{equation}
From  (\ref{eq:SAdSM}) and (\ref{eq:EMTMAdS}) we find the energy-momentum tensor of the
matter field $\Phi$
\begin{equation} \label{eq:EMs}
\Theta_{\ell m} \! = \partial_\ell \Phi^* \partial_m \Phi + \partial_m \Phi^* \partial_\ell \Phi 
- g_{\ell m} \! \left(\partial^n \Phi^* \partial_n \Phi - \! \mu^2 \Phi^* \Phi\right).
\end{equation}

Likewise, we can determine the AdS equation of motion of the graviton field $h_{\ell m}$   by substituting
the modified metric $\bar g_{\ell m} = g_{\ell m} + h_{\ell m}$ into the gravitational action
$S_G$. We find 
\begin{multline} \label{eq:SGh}
S_G[h_{\ell m}] = S_G[0]  \\ + \frac{1}{4 \kappa^2} \!
\int \! d^{d+1}x  \, \sqrt{g} \, \Big(\partial_n h^{\ell m} \partial^n h_{\ell m} 
- \half \partial_\ell h \, \partial^\ell h\Big)  + \mathcal{O}(h^2),
\end{multline}
where  the trace $h_\ell^\ell$ is denoted by $h$.  In deriving (\ref{eq:SGh}) we have made use of the gauge invariance of the theory $h'_{\ell m} = h_{\ell m} +
\partial_\ell \epsilon_m + \partial_m \epsilon_\ell$ to impose 
the harmonic gauge condition $\partial_\ell h^\ell_m = \half \partial_m h$. The action describing the dynamical fields $h_{\ell m}$ in the
weak field approximation ($d=4$)  is given in the linearized form
\begin{equation}  \label{eq:Sh}
S_h = \frac{1}{4 \kappa^2} \!
\int \! d^4x \, dz   \sqrt{g} \Big(\partial_n h^{\ell m} \partial^n h_{\ell m} 
- \half \partial_\ell h \, \partial^\ell h\Big),
\end{equation}
resembling the treatment of an ordinary  gauge field.
The total bulk action describing the coupling of gravity and matter with an external
graviton in the weak field approximation thus has two additional terms: $S = S_G +S_M + S_h + S_I$.

\section{Hadronic States and Transition Matrix Elements in AdS/CFT}
\label{TME-AdS}

A physical hadron in four-dimensional Minkowski space has four-momentum $P_\mu$ and invariant
hadronic mass states determined by the light-front eigenvalue equation
$H_{LF} \vert \psi_P \rangle = \mathcal{M}^2 \vert \psi_P \rangle$.
On AdS space the physical  states are
represented by normalizable modes
\begin{equation} \label{eq:PhiP}
\Phi_P(x,z) = e^{-iP \cdot x} \Phi(z),
\end{equation}
with plane waves along the Poncar\'e coordinates and a profile function $\Phi(z)$ 
along the holographic coordinate $z$. The hadronic invariant mass 
$P_\mu P^\mu = \mathcal{M}^2$ for the string 
modes (\ref{eq:PhiP}) is found by solving the eigenvalue problem for the
AdS wave equation. Each  light-front hadronic state $\vert \psi_P \rangle$ is dual to a normalizable
string mode $\Phi_P(x,z)$. To compare a physical observable computed in light-front QCD
with the same observable computed in AdS space, we must find a gauge-invariant 
prescription to relate physical states in both theories. In practice, one compares
the results of matrix elements of local operators on the gauge theory side with
the corresponding matrix element in the AdS side. A consistent normalization
on both sides of the correspondence is  determined by the normalization
of hadronic states to the energy-momentum tensor. 

\subsection{Normalization of Hadronic States to the Energy-Momentum Tensor in AdS}

We compute the expectation value of the energy momentum tensor $\Theta_{\ell m}$ along
Minkowski coordinates. For $d=4$
\begin{equation}
\left\langle \Phi_P \left\vert \Theta_\mu^{\, \nu} \right\vert \Phi_P \right\rangle =
\int \! d^4x \,dz  \sqrt{g} \,\Theta_\mu^{\, \nu} .
\end{equation}
Substituting the plane-wave solution (\ref{eq:PhiP}) in the expression for the energy-momentum 
tensor (\ref{eq:EMs}) we find
\begin{equation} \label{eq:EMTz}
\left\langle P \left\vert \Theta_\mu^{\, \nu} \right\vert P \right\rangle =
2 P_\mu P^\nu,
\end{equation}
where we have extracted the overall factor  $(2 \pi)^4 \delta^{(4)} \! \left(P' - P\right) $ from the $x$-integration to compare with the light-front QCD results. 
We chose the normalization
\begin{equation}
R^3 \! \int_0^{\Lambda_{\rm QCD}^{-1}} \! \frac{dz}{z^3}  \, \left\vert \Phi(z) \right\vert^2 = 1,
\end{equation}
in the cutoff AdS space,  consistent with the
normalization of AdS solutions to the total charge operator~\cite{Brodsky:2007hb}
described in Appendix \ref{JAdS}. In obtaining (\ref{eq:EMTz}) we have dropped the last term in (\ref{eq:EMs}), a surface term which vanishes by choosing appropriate boundary conditions.

\subsection{Hadronic Transition Matrix Elements in AdS/CFT}

The matrix element of the energy-momentum tensor  for the hadronic transition
$P \to P'$,
follows from the interaction term  (\ref{eq:SI})  describing the
coupling of the pion mode with the external graviton field propagating in AdS space
\begin{equation}
 \int \! d^4x \, dz \sqrt{g}\, h_{\ell m}  \left(
\partial^{\ell} \Phi_{P'}^* \partial^{m} \Phi_P+
\partial^{m} \Phi_{P'}^* \partial^{\ell} \Phi_P \right),
\label{eq:T}
\end{equation} 
where we have dropped the surface term in (\ref{eq:EMs}).

Since the energy momentum tensor $\Theta^{\ell m}$ is gauge invariant, we may impose a more
restricted gauge condition in order to simplify the calculations and use the general covariance of the theory to obtain the final result. We choose the harmonic-traceless gauge
 $\partial_\ell h^\ell_m = \half \partial_m h = 0$ and
we consider the propagation inside AdS space of a graviton probe  $h_{\ell m}$ with
metric components
along Minkowski coordinates $h_{zz} = h_{z \mu} = 0$.  
The set of linearized Einstein equations from (\ref{eq:Sh}) reduce to the simple form~\cite{Abidin:2008ku}
\begin{equation} \label{eq:AdSh}
z^3 \partial_z \left( \frac{1}{z^3} \partial_z h_\mu^{\, \nu} \right)
- \partial_\rho  \partial^\rho h_\mu^{\, \nu} = 0.
\end{equation}

To solve (\ref{eq:AdSh}) we note that the boundary limit of the graviton probe is a plane wave
along the Poincar\'e coordinates with polarization indices  along the physical transverse
dimensions $h_\mu^{\, \nu}(x, z \to 0) = \epsilon_\mu^{\, \nu} e^{- i q \cdot x}$,
where $q^2 = - Q^2 < 0$. 
As discussed in~\cite{Abidin:2008ku}  in this particular gauge, the graviton couples to
the transverse and traceless part of the energy-momentum tensor. 
We thus write
\begin{equation} \label{eq:hmunu}
h_\mu^{\, \nu}(x, z ) = \epsilon_\mu^{\, \nu} \, e^{- i q \cdot x} H(q^2, z),
\end{equation}
with
\begin{equation} \label{eq:Hbc}
H(q^2= 0, z) = H(q^2, z = 0) = 1.
\end{equation}
Substituting $h_\mu^\nu$ in (\ref{eq:AdSh}) we find the wave equation describing
the propagation of the external graviton inside AdS space
\begin{equation} \label{eq:AdSJ}
\left[ z^2 \partial_z^2 -  3 z \, \partial_z - z^2 Q^2 \right]   H(Q^2, z)  = 0.
 \end{equation}
Its solution subject to the boundary conditions (\ref{eq:Hbc}) is
\begin{equation} \label{eq:H}
H(Q^2, z) = \half  Q^2 z^2  K_2(z Q),
\end{equation}
the result obtained by Abidin and Carlson~\cite{Abidin:2008ku}.

We can now use the Minkowski space dependence of the normalizable mode 
$\Phi_P(x,z) = e^{-iP \cdot x} \Phi(z)$ in (\ref{eq:T}).
We find the transition amplitude
\begin{equation}
\left\langle P' \left\vert \Theta_\mu^{\, \nu} \right\vert P \right\rangle 
=  \left( P^\nu P'_\mu + P_\mu  P'^\nu \right) A(Q^2),
\end{equation}
where we have extracted the overall factor $(2 \pi)^4 \delta^{(4)} \! \left(P' - P - q\right)$ from 
momentum conservation at the vertex from integration over Minkowski variables.
We find for $A(Q^2)$
\begin{equation} 
A(Q^2)  =   R^3 \! \! \int \frac{dz}{z^3} \, \Phi(z) H(Q^2, z) \Phi(z),
\label{eq:AdSA}
\end{equation}
with $A(Q^2 \! = \! 0) = 1$.
The gravitational form-factor in AdS is thus 
represented as the $z$-overlap
of the normalizable modes dual to the incoming
and outgoing hadrons, $\Phi_P$ and $\Phi_{P'}$, with the
non-normalizable mode, $H(Q^2, z)$, dual to the external
graviton source~\cite{Abidin:2008ku}; this provides the form of the 
gravitational transition matrix element analogous to the electromagnetic form-factor
in AdS~\cite{Polchinski:2002jw}.
At small $z$ the string modes
scale as $\Phi \sim z^\Delta$.
At large enough $Q$, the important contribution to (\ref{eq:AdSA})
is from the region near $z \sim 1/Q$, 
$A(Q^2) \to \left(1/Q^2\right)^{\Delta - 1}$,
and the ultraviolet point-like behavior responsible for the power law scaling~\cite{Brodsky:1973kr, Matveev:ra} is recovered. 

\section{Light-Front Mapping of String Amplitudes}
\label{LFM}

The gravitational form factor (\ref{eq:AdSA})
represents the coupling of the graviton to the entire hadron in AdS, independent of the number $n$ of its constituents.  Since  (\ref{eq:AdSA}) gives $A(0) =1$, it implicitly includes the sum over the coupling of the graviton to all $n$ massless constituents. The gravitational coupling, like a number operator, sums over all particles. Similarly, the electromagnetic current sums over constituents, but weighted by their fractional charge. To simplify the discussion we will establish the connection of the AdS/CFT results for the gravitational form factor and the light-front results for the lowest Fock state $n=2$ using the effective single particle distribution discussed in Sec. \ref{ESPD}. This is particularly useful for extending the
results to  arbitrary $n$, subject to the requirement that  one normalizes the hadronic matrix element of the energy momentum tensor so that $A(0)=1$.

For $n=2$, there are two terms which contribute to the light-front result in the $f$-sum in 
(\ref{eq:GFFzeta}). 
Exchanging $x \leftrightarrow 1-x$ in the second integral we find
\begin{multline}  \label{eq:PiGFF} 
A_{n=2}(q^2) =  4 \pi \! \int_0^1 \! dx \, (1-x)  \\ \times \int \zeta d \zeta 
J_0 \! \left(\! \zeta q \sqrt{\frac{1-x}{x}}\right)  \big\vert \tilde\rho_{n=2} (x,\zeta) \big\vert^2,
\end{multline}
where $\zeta^2 =  x(1-x) \mathbf{b}_\perp^2$.  It is simple to prove that if 
$\tilde\rho$ is a symmetric function of $x$ and  $1-x$ then
\begin{equation}
2 \pi \!\int_0^1 \! dx \, (1-x)  \int \! \zeta \, d \zeta  \,
\big\vert \tilde \rho_{n=2}(x, \zeta) \big\vert^2 = \frac{1}{2},
\end{equation}
and thus $A(q^2)$ satisfies the sum rule $A(0) = 1$.

To compare with the light-front QCD results we express the bulk-to-boundary 
propagator $H(Q^2, z)$ (\ref{eq:H}) for the graviton probe
using the Hankel-Nicholson integral representation (Appendix A of Reference \cite{Brodsky:2007hb}) 
\begin{equation}
H(Q^2,z) = 4 Q^4 
\int_0^\infty \frac{t J_0(z t)}
 {\left(t^2 + Q^2\right)^3} \, dt  .
\end{equation}
Introducing a new variable $x = \frac{Q^2}{t^2 + Q^2}$ we find
\begin{equation} \label{eq:intHz}
H(Q^2, z) =  2 \int_0^1\!  x \, dx \, J_0\!\left(\!z Q\sqrt{\frac{1-x}{x}}\right) ,
\end{equation}
and thus
\begin{equation} 
A(Q^2)  =   2 R^3 \! \int_0^1 \! x \, dx  \! \int \frac{dz}{z^3} \, 
J_0\!\left(\!z Q\sqrt{\frac{1-x}{x}}\right) \left \vert\Phi(z) \right\vert^2 .
\label{eq:AdSAx}
\end{equation}

We can now compare the above expression with the light-front expression (\ref{eq:PiGFF}),  and can identify the spectator density
function appearing in the light-front 
formalism with the corresponding AdS density
\begin{equation} \label{eq:PhirhoHW}
\tilde \rho(x,\zeta)
=    \frac{R^3}{2 \pi} \frac{x}{1-x}
\frac{\left\vert \Phi(\zeta)\right\vert^2}{\zeta^4} .
\end{equation}
Extension to arbitrary $n$ follows from the $x$-weighted definition of the transverse impact
variable of the $n-1$ spectator system given by (\ref{eq:zeta}): 
$\zeta = \sqrt{\frac{x}{1-x}} \, \Big\vert \sum_{j=1}^{n-1} x_j \mathbf{b}_{\perp j}\Big\vert$.

Equation (\ref{eq:PhirhoHW}) holds for all momentum transfer $Q^2$ and gives the same relation between  string modes $\Phi(\zeta)$ 
in AdS$_5$ and the QCD transverse charge density $\tilde\rho(x,\zeta)$  obtained
previously
by mapping the electromagnetic current matrix elements~\cite{Brodsky:2007hb}. 
The variable $\zeta$, $0 \leq \zeta \leq \Lambda_{\rm QCD}^{-1}$, represents a measure of the transverse separation between point-like constituents, and it is also the
holographic variable $z$.

In the case of a two-parton system the correspondence between the string 
amplitude $\Phi(z)$ in AdS space and the QCD light-front  wavefunction 
 $\tilde \psi\left(x, \mbf{b}_\perp\negthinspace\right)$
follows from (\ref{eq:PhirhoHW}). For two partons the transverse density
(\ref{eq:rhoeta}) has the simple form
\begin{equation}
\tilde\rho_{n=2}(x, \zeta) = 
\frac{\vert \tilde\psi(x,\zeta)\vert^2}{(1-x)^2},
\end{equation}
and a closed form solution for the two-constituent bound state
light-front wave function is obtained
\begin{equation} \label{eq:Phipsi}
\vert \tilde\psi(x,\zeta)\vert^2 = 
\frac{R^3}{2 \pi} \, x(1-x)
\frac{\vert \Phi(\zeta)\vert^2}{\zeta^4},
\end{equation}
with $\zeta^2 =  x(1-x) \mathbf{b}_\perp^2$.
For a two-parton system the light-front mapping can also be carried out directly from (\ref{eq:MGFFbb}).
This is done in Appendix \ref{PionLFM}, where the consistency with the LF mapping results from electromagnetic current matrix elements is also pointed out.

\subsection{Holographic Light-Front Hamiltonian and Schr\"odinger Equation}

The above analysis provides an exact correspondence between the holographic variable $z$ and an
impact variable $\zeta$ which measures the transverse separation between point-like constituents within
a hadron; we can identify $\zeta = z$. 
The mapping of $z$ from AdS space to $\zeta$ in light-front frame  allows the equations of motion in AdS space to be recast in the form of  a
light-front Hamiltonian equation~\cite{Brodsky:1997de} with eigenvalues given in terms
of the hadronic eigenmass $\mathcal{M}^2$
\begin{equation}
H_{LF} \ket{\phi} = \mathcal{M}^2 \ket{\phi}, \label{eq:HLC}
\end{equation}
a remarkable result which allows the discussion of the AdS/CFT solutions in terms of light-front equations in physical 3+1 space time.

Factoring out the plane wave dependence of the hadronic mode
$\Phi_P(x,z) = e^{-iP \cdot x} \Phi(z)$ and substituting in (\ref{eq:AdSPhi})
we find 
\begin{equation} \label{eq:AdSPhiz}
\left[ z^2 \partial_z^2 -  (d-1) z \, \partial_z + z^2 \mathcal{M}^2  - (\mu R)^2\right]   \Phi(z)  = 0,
 \end{equation}
the wave equation describing the propagation of a scalar mode in AdS.
The allowed values of $\mu$ are determined by requiring that asymptotically the
dimensions become separated by integers according to the spectral relation
$(\mu R)^2 = \Delta(\Delta - d)$ and the stability condition dictated by the Breitenlohner-Freedman bound 
$(\mu R)^2 \ge - d^2/4$ for a scalar field~\cite{Breitenlohner:1982jf}.
We find $(\mu R)^2 = - 4 + L^2$ 
for $\Delta = 2+L$ and $d=4$. Thus the stability bound requires $L^2 \ge 0$.

By substituting
\begin{equation}
\phi(\zeta) = \left(\frac{\zeta}{R}\right)^{-3/2} \Phi(\zeta)
\end{equation} in the AdS scalar wave equation (\ref{eq:AdSPhiz})
we find an effective Schr\"odinger equation as a function of the
weighted impact variable $\zeta$~\cite{Brodsky:2006uqa,Brodsky:2007hb}
\begin{equation} \label{eq:Scheq}
\left[-\frac{d^2}{d \zeta^2} + V(\zeta) \right] \phi(\zeta) =
\mathcal{M}^2 \phi(\zeta),
\end{equation}
with $-\frac{d^2}{d \zeta^2}$ the light-front kinetic energy operator and
conformal potential
\begin{equation}
V(\zeta) \to - \frac{1-4 L^2}{4\zeta^2},
\end{equation}
an effective relativistic two-particle light-front wave equation for mesons
defined at $x^+ = 0$.
Its eigenmodes determine the hadronic mass spectrum. 

In the transverse impact holographic representation with holographic
light-front wavefunctions $\phi(\zeta) = \langle \zeta \vert \phi \rangle$,
the LC eigenvalue equation thus reads
\begin{equation}
\langle \zeta \vert H_{LC} \vert \phi \rangle =
\mathcal{M}^2  \langle \zeta \vert \phi \rangle,
\end{equation}
with
\begin{equation}
\langle \zeta \vert H_{LC} \vert \phi \rangle =
\left[ -\frac{d^2}{d \zeta^2} 
-  \frac{1-4 \nu^2}{4\zeta^2} \right] 
\langle \zeta \vert \phi \rangle,
\label{eq:HLCzeta}
\end{equation}
in the conformal limit. 
The light-front modes $\phi(\zeta) = \langle \zeta \vert \phi \rangle$ are
normalized according to
\begin{equation}
\langle \phi \vert \phi \rangle = \int d\zeta \, \vert \langle
\zeta \vert \phi \rangle \vert^2 = 1,
\end{equation}
and represent the probability amplitude to find $n$-partons at
transverse impact separation $\zeta = z$.   Its
eigenvalues are determined by the boundary conditions 
$\phi(z =1/\Lambda_{\rm QCD}) = 0$ and are given in terms of the roots of
Bessel functions: $\mathcal{M}_{L,k} = \beta_{L,k} \Lambda_{\rm
QCD}$. The normalizable modes are
\begin{equation}
\phi_{L,k}( \zeta) =   \frac{ \sqrt{2} \Lambda_{\rm QCD}}{J_{1+L}(\beta_{L,k})}
 \sqrt{\zeta} J_L \! \left(\zeta \beta_{L,k} \Lambda_{\rm QCD}\right)
 \theta\big(\zeta \le
\Lambda^{-1}_{\rm QCD}\big).
\end{equation}

The lowest stable state $L = 0$ is determined by the
Breitenlohner-Freedman bound.
Higher excitations are matched to the small $z$ asymptotic behavior of each string mode
to the corresponding
conformal dimension of the boundary operators
of each hadronic state. The AdS metric $ds^2$ (\ref{eq:AdSzLF})  is invariant if
$\mbf{x}_\perp^2 \to \lambda^2 \mbf{x}_\perp^2$ and $z \to \lambda
z$ at equal light-front time  $x^+ = 0$. The effective wave equation (\ref{eq:Scheq}) has the Casimir representation $L^2$ corresponding
to the $SO(2)$ group of rotations in the transverse light-front plane.
Indeed, the Casimir operator for $SO(N) \sim S^{N-1}$ is $L(L+N -2)$.
This shows the natural holographic connection to the light front.
The fundamental light-front equation of AdS/CFT has the appearance of a
Schr\"odinger  equation, but it is relativistic, covariant, and analytically tractable.

A closed form of the light-front wavefunctions $\tilde\psi(x, \mbf{b}_\perp)$ for a two-parton
bound state follows from
(\ref{eq:Phipsi}) 
\begin{multline} 
\tilde \psi_{L,k}(x, \mbf{b}_\perp) 
=  \frac {\Lambda_{\rm QCD}}{\sqrt{ \pi} J_{1+L}(\beta_{L,k})} \sqrt{x(1-x)} \\ \times
J_L \! \left(\sqrt{x(1-x)} \, \vert\mbf{b}_\perp\vert \beta_{L,k} \Lambda_{\rm QCD}\right) 
\theta \! \left(\mbf{b}_\perp^2 \le \frac{\Lambda^{-2}_{\rm QCD}}{x(1-x)}\right).
\end{multline}

\begin{figure*}[htp]
\centering
\includegraphics[scale=0.85]{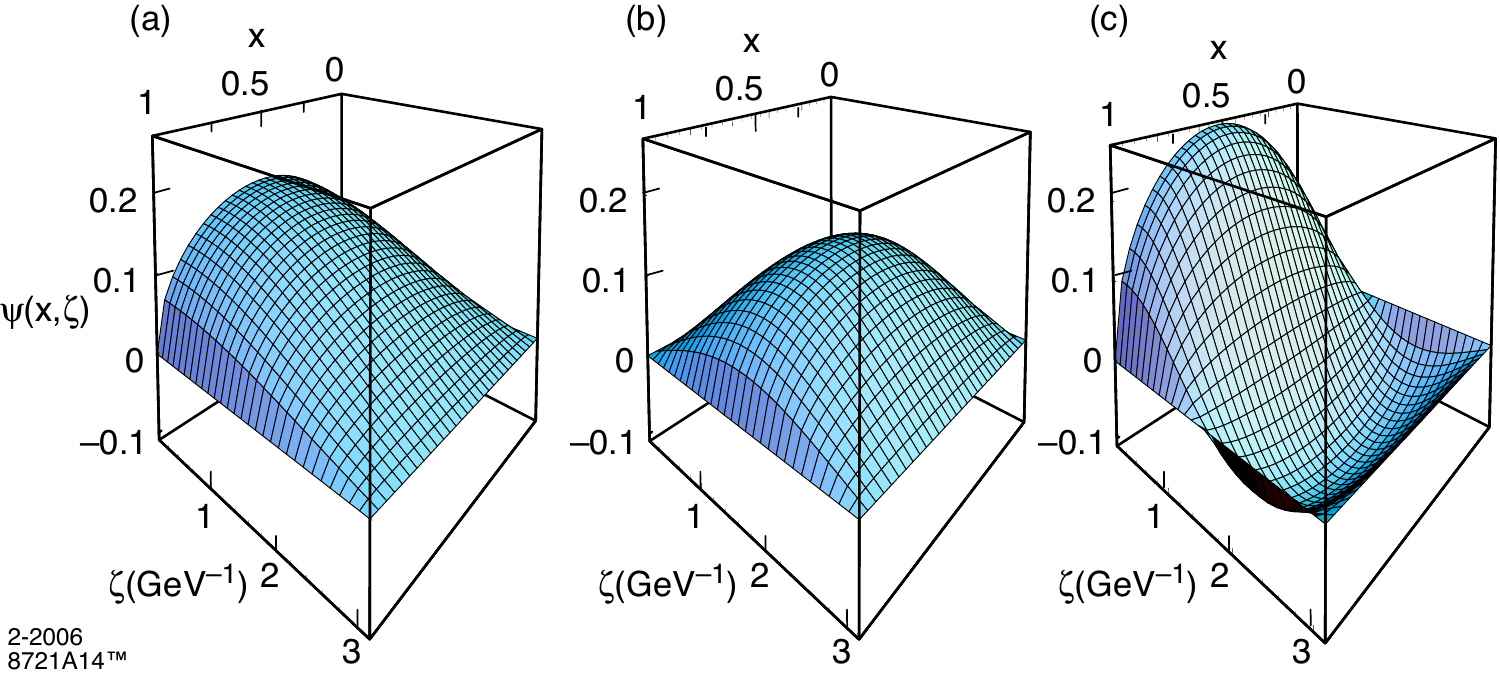}
\caption{AdS/QCD Predictions for the light-front wave functions of a
meson in the hard-wall model: (a) $n = 0, L = 0$; (b) $n = 0, L = 1$; (c)  $n = 1, L = 0$.}
\label{fig:HWLFWF}
\end{figure*}

The resulting wavefunction depicted in Fig. \ref{fig:HWLFWF} 
displays confinement at large interquark
separation and conformal symmetry at short distances, reproducing dimensional counting rules for hard exclusive amplitudes and the conformal properties of the LFWFs at high relative
momenta~\cite{Brodsky:1973kr, Matveev:ra}.

For the soft-wall model~\cite{Karch:2006pv} we can obtain the basis set of light-front wavefunctions by comparing the QCD expression for the gravitational form factor (\ref{eq:GFFzeta}) 
with the corresponding expression for the AdS form factor, where the graviton probe propagates in the
distorted metric.  In the large $Q$ limit we can identify the light-front
transverse density with the corresponding AdS density, with identical results as obtained for the mapping of the electromagnetic form factor in~\cite{Brodsky:2007hb}. 

\section{Conclusions}
\label{Conclusions}

Light-Front Holography is one of the most remarkable features of AdS/CFT.  It  allows one to project the functional dependence of the wave function $\Phi(z)$ computed  in the single AdS fifth dimension to the  hadronic frame-independent light-front wave function $\psi(x_i, \mbf{b}_{\perp i})$ in $3+1$ physical space-time. The 
variable $z $ maps  to $ \zeta(x_i, \mbf{b}_{\perp i})$.   As we have discussed, this correspondence is a consequence of the fact that the metric $ds^2$ for AdS$_5$ at fixed light-front time $\tau$ is invariant under the simultaneous scale change  $\mbf{x}^2_\perp \to \lambda^2 \mbf{x}^2_\perp $ in transverse space and $z^2 \to \lambda^2 z^2$.  The transverse coordinate $\zeta$ is closely related to the invariant mass squared  of the constituents in the LFWF  and its off-shellness  in  the light-front kinetic energy,  and it is thus the natural variable to characterize the hadronic wavefunction.  In fact $\zeta$ is the only variable to appear in the light-front
Schr\"odinger equations predicted from AdS/QCD.  These equations for both meson and baryons give a good representation of the observed hadronic spectrum, especially in the case of the soft wall model. The resulting LFWFs also have excellent phenomenological features, including predictions for the  electromagnetic form factors and decay constants.  

It is interesting to note that the form of the nonperturbative pion distribution amplitude $ \phi_\pi(x)$ obtained from integrating the $ q \bar q$ valence LFWF $\psi(x, \mbf{k}_\perp)$  over $\mbf{k}_\perp$,
has a quite different $x$-behavior than the
asymptotic distribution amplitude predicted from the PQCD
evolution~\cite{Lepage:1979zb} of the pion distribution amplitude.
The AdS prediction
$ \phi_\pi(x)  = \sqrt{3}  f_\pi \sqrt{x(1-x)}$ has a broader distribution than expected from solving the Efremov-Radyushkin-Brodsky-Lepage (ERBL) evolution equation in perturbative QCD.
This observation appears to be consistent with the results of the Fermilab diffractive dijet experiment~\cite{Aitala:2000hb}, the moments obtained from lattice QCD~\cite{Brodsky:2008pg}, and pion form factor 
data~\cite{Choi:2006ha}.

Nonzero quark masses are naturally incorporated into the AdS predictions~\cite{Brodsky:2008pg} by including them explicitly in the LF kinetic energy  $\sum_i  \frac{\mbf{k}^2_{\perp i} + m_i^2}{x_i}$.  Given the nonpertubative LFWFs one can predict many interesting phenomenological quantities such as heavy quark decays, generalized parton distributions and parton structure functions.  
The AdS/QCD model is semiclassical and thus only predicts the lowest valence Fock-state structure of the hadron LFWF. In principle, the model can be systematically improved by diagonalizing the full QCD light-front Hamiltonian on the AdS/QCD basis.

Another interesting application is hadronization at the amplitude level.  In this case one uses light-front time-ordered perturbation theory for the QCD light-front Hamiltonian to generate the off-shell  quark and gluon T-matrix helicity amplitudes such as $e^+ e^- \to \gamma^* \to X$.  If at any stage a set of  color-singlet partons has  light-front kinetic energy 
$\sum_i {\mbf{k}^2_{\perp i}/ x_i} < \Lambda^2_{QCD}$, then one coalesces the virtual partons into a hadron state using the AdS/QCD LFWFs. A similar approach was used to predict antihydrogen formation from virtual positron--antiproton states produced in $\bar p A$ collisions~\cite{Munger:1993kq}.

The hard-wall AdS/QCD model resembles  bag models where a boundary condition is introduced to implement confinement.  However, unlike traditional bag models, the AdS/QCD model is frame-independent.  An important property of bag models is the  dominance of quark interchange as the underlying dynamics of large-angle elastic scattering.  This agrees with the survey of 
two-hadron exclusive reactions~\cite{White:1994tj}.  In addition the AdS/QCD model implies a maximal wavelength for confined quarks and gluons and thus a finite IR fixed point for the QCD coupling.

We originally derived the light-front holographic mapping by matching the exact expression for current matrix elements in AdS space with the corresponding exact expression for the electromagnetic  current matrix element using light-front theory in physical space-time.  In this paper we have shown that one obtains the identical holographic mapping using the hadronic matrix elements of the energy-momentum tensor. This is a highly nontrivial test of the consistency of the light-front holographic mapping.

Our analysis also allows one to predict the individual quark and gluon contributions to the gravitational form factors $A(q^2)$ and $B(q^2).$  Thus we can predict the momentum fractions 
for quarks $q$ and gluons $g$, $A_{q,g}(0) =  \langle x_{q,g}\rangle$, and orbital angular momenta, $B_{q,g}(0) = \langle L_{q,g} \rangle$, carried by each quark flavor and gluon in the hadron with sum rules $\sum_{q,g}A_{q,g}(0) = A(0)= 1$ and $\sum_{q,g} B_{q,g}(0) =B(0) = 0$.  The last sum rule corresponds to the vanishing of the  anomalous gravitational moment which is true Fock state by 
Fock state~\cite{Brodsky:2000ii} in light-front theory. 

The mathematical consistency of light-front holography for both the electromagnetic and gravitational hadronic transition matrix elements demonstrates that the mapping between the single AdS space dimension $z$ and the transverse light-front variable $\zeta,$ which is a function of the multi-dimensional coordinates of the partons in a given light-front Fock state $x_i, \mbf{b}_{\perp i}$ at fixed light-front time $\tau,$ is a general principle. The holographic mapping from $\Phi(z)$ to the light-front wave functions of relativistic composite systems provides a new tool for extending the AdS/CFT correspondence to theories such as QCD which are not conformally invariant.

\begin{acknowledgments}

This work was completed at the Institute for Nuclear Theory at the University of Washington at Seattle
during the workshop 
``From Strings to Things: String Theory Methods in QCD and Hadron Physics".
The authors would 
like to thank the members of the Institute for their hospitality.
SJB thanks the C.N. Yang Institute for Theoretical Physics at the State University of New York and the High Energy Theory Group at Brookhaven National Laboratory for their hospitality. 
GFdT thanks the Centre de Physique Th\'eorique at Ecole Polytechnique and the
Laboratoire de Physique Th\'eorique at Ecole Normale Sup\'erieure in Paris for
their hospitality.
We also thank Professor Robert Shrock for helpful conversations.
This research was supported by the Department
of Energy contract DE--AC02--76SF00515. 

\end{acknowledgments}

\appendix

\section{Metric Conventions}
\label{Metric}

\subsection{Light-Cone Metric and Minkowski Space}

The Minkowski metric written in terms of light front coordinates is
\begin{equation}
d\sigma^2 = dx^+ dx^- - d \mbf{x}_\perp^2 - dz^2,
\end{equation}
with timelike and spacelike components $x^+ = x^0 + x^3$ and $ x^- = x^0 - x^3$ respectively.
We write  contravariant four-vectors such as $x^\mu$ as
\begin{equation}
x^\mu = \left(x^+, x^-, x^1, x^2\right) = \left(x^+, x^-, \mbf{x}_\perp\right).
\end{equation}
Scalar products are
\begin{eqnarray} \nonumber
x \cdot p &\! = \!&  x_\mu p^\nu = {\rm g}_{\mu \nu} x^\mu p^\nu \\ \nonumber
&\! = \!& x_+p^+ + x_-p^- + x_1 p^1 + x_2 p^2   \\
& \!= \!& \frac{1}{2} \left(x^+ p^- + x^- p^+\right) - \mbf{x}_\perp \cdot \mbf{p}_\perp,
\end{eqnarray}
with front-form metrics
\begin{equation}
{\rm g}_{\mu \nu} =
  \begin{pmatrix}
  0 &  \frac{1}{2} & 0 & 0 \\
  \tfrac{1}{2} & 0  & 0 & 0 \\
  0 & 0  & -1 & 0 \\
  0 & 0  & 0 & -1 \\
  \end{pmatrix} ,
  \end{equation}
  and
  \begin{equation}
  {\rm g}^{\mu \nu} =
  \begin{pmatrix}
  0 &  2 & 0 & 0 \\
  2 & 0  & 0 & 0 \\
  0 & 0  & -1 & 0 \\
  0 & 0  & 0 & -1 \\
  \end{pmatrix}.
  \end{equation}
  A covariant vector such as $\partial_\mu$ is
  \begin{equation}
  \partial_\mu = \left(\partial_+, \partial_-, \partial_1, \partial_2\right)=
  \left(\partial_+, \partial_-, \vec \partial_\perp\right).
  \end{equation}
  Thus $\partial^+ = 2 \, \partial_-$ and $\partial ^- = 2 \, \partial_+$.
  
 \subsection{AdS Space}
 
 AdS coordinates are the Minkowski coordinates $x^\ell$ and $z$ labelled
 $x^\ell = \left(x^\ell, z\right)$. The AdS$_{d+1}$ metric is 
 \begin{eqnarray} \label{eq:gAdS}
ds^2 &\! = \! & g_{\ell m} dx^\ell dx^m \nonumber \\
&\!=\!& \frac{R^2}{z^2} \left(\eta_{\mu \nu} dx^\mu dx^\nu - dz^2\right),
 \end{eqnarray}
 with conformal metrics
 \begin{equation}
 g_{\ell m} = \frac{R^2}{z^2}
  \begin{pmatrix}
  1 &  0 & 0 & 0 \\
  0 & -1  & 0 & 0 \\
  0 & 0  & \ddots & 0 \\
  0 & 0  & 0 & -1 \\
  \end{pmatrix} ,
   \end{equation}
  and
  \begin{equation}
  g^{\ell m} = \frac{z^2}{R^2}
  \begin{pmatrix}
  1 &  0 & 0 & 0 \\
  0 & -1  & 0 & 0 \\
  0 & 0  & \ddots & 0 \\
  0 & 0  & 0 & -1 \\
  \end{pmatrix}.
  \end{equation}
The AdS metric is conveniently written $g_{\ell m} = \frac{R^2}{z^2} \eta_{\ell m}$ and
$g^{\ell m} = \frac{z^2}{R^2} \eta^{\ell m}$, where $\eta_{\ell m}$ has diagonal components
$(1, -1, \cdots, -1)$. The metric determinant $g = \left\vert g_{\ell m} \right\vert$ is
$g = \left(\frac{R^2}{z^2}\right)^{d+1}$.
  
 \section{Normalization of Hadronic States to the Charge Operator in AdS}
 \label{JAdS}
 
We compute the expectation value of the electromagnetic current $J_\ell$ 
\begin{equation}
J_\ell = i \left( \Phi^* \partial_\ell \Phi - \Phi \partial_\ell \Phi^* \right),
\end{equation}
along Minkowski coordinates for a
hadronic state $\Phi_P(x,z) = e^{-iP \cdot x} \Phi(z)$ in AdS$_5$ space
\begin{equation} 
\left\langle \Phi_P \left\vert J^\mu \right\vert \Phi_P \right\rangle =
\int \! d^4x \,dz  \sqrt{g} \, J^\mu .
\end{equation}
Substituting the hadronic plane-wave solution we obtain
\begin{equation} \label{eq:JPhi}
\left\langle P \left\vert J^\mu \right\vert P \right\rangle =
2 P^\mu,
\end{equation}
where we have extracted the overall factor  $(2 \pi)^4 \delta^{(4)} \! \left(P' - P\right) $ from the $x$-integration to compare with the light-front QCD results described in \cite{Brodsky:2007hb}. We  use the normalization
\begin{equation}
R^3 \! \int_0^{\Lambda_{\rm QCD}^{-1}} \! \frac{dz}{z^3}  \, \left\vert \Phi(z) \right\vert^2 = 1,
\end{equation}
in the cutoff AdS space.
The total charge operator is a diagonal operator in the AdS hadronic representation.

 \section{A Two-Parton Example}
  \label{PionLFM}
 
 The mapping of AdS transition amplitudes to light-front QCD transition matrix elements
 is much simplified for two-parton  hadronic states. It further illustrates  important technical aspects
 for extending the results to the $n$-parton case. We describe
 in this appendix the actual two-parton mapping for the electromagnetic and gravitational transition amplitudes.
 
 \subsection{Electromagnetic Form Factor}

The Drell-Yan-West expression for the electromagnetic form-factor in impact 
space~\cite{Brodsky:2007hb, Brodsky:2006uqa}
\begin{multline} \label{eq:FFb} 
F(q^2) =  \sum_n  \prod_{j=1}^{n-1}\int d x_j d^2 \mbf{b}_{\perp j}  \\\sum_q e_q
\exp \! {\Bigl(i \mbf{q}_\perp \! \cdot  \! \sum_{k=1}^{n-1} x_k \mbf{b}_{\perp k}\Bigr)} 
\left\vert \tilde \psi_n(x_j, \mbf{b}_{\perp j})\right\vert^2,
\end{multline}
is written as a sum of overlap integrals of light-front wave functions of the $j = 1,2, \cdots, n-1$ spectator
constituents. We have included
explicitly in (\ref{eq:FFb})
the contribution from each active constituent $q$ with charge $e_q$.
The formula is exact if the sum is over all Fock states $n$.

For definiteness we shall consider a two-quark $\pi^+$  valence Fock state 
$\vert u \bar d\rangle$ with charges $e_u = \frac{2}{3}$ and $e_{\bar d} = \frac{1}{3}$.
For $n=2$, there are two terms which contribute to the $q$-sum in (\ref{eq:FFb}). 
Exchanging $x \leftrightarrow 1-x$ in the second integral  we find ($e_u + e_{\bar d}$ = 1)
\begin{equation} 
F_{\pi^+}(q^2)  =  \!  \int_0^1 \! d x \int \! d^2 \mbf{b}_{\perp}  
 e^{i \mbf{q}_\perp \cdot  \mbf{b}_{\perp} (1-x)} 
\left\vert \tilde \psi_{u \bar d/ \pi}\! \left(x,  \mbf{b}_{\perp }\right)\right\vert^2 ,
\end{equation}
with normalization $F_\pi^+(q\!=\!0)=1$. Integrating over angle we find
\begin{multline} \label{eq:PiFFb} 
F_{\pi^+}(q^2) =  2 \pi \int_0^1 \! \frac{dx}{x(1-x)}  \\ \times \int \zeta d \zeta  
  J_0 \! \left(\! \zeta q \sqrt{\frac{1-x}{x}}\right) 
\left\vert\tilde\psi_{u \bar d/ \pi}\!(x,\zeta)\right\vert^2,
\end{multline}
where $\zeta^2 =  x(1-x) \mathbf{b}_\perp^2$.
Notice that by performing an identical calculation for the
$\pi^0$ meson the result is $F_{\pi^0}(q^2) = 0$ for any $q$, as expected
from $C$-charge conjugation invariance.

We now compare this result with the electromagnetic form-factor in AdS space~\cite{Brodsky:2007hb, Brodsky:2006uqa}:
\begin{equation} 
F(Q^2) = R^3 \int \frac{dz}{z^3} \, J(Q^2, z) \left\vert \Phi_{\pi^+}(z) \right\vert^2,
\label{eq:FFAdS}
\end{equation}
where $F(Q^2\!=\!0) = 1$ and the bulk-to-boundary propagator $J(Q^2, z) = z Q K_1(z Q)$ 
describes the propagation of the external electromagnetic current inside AdS.
Using the integral representation  of $J(Q^2,z)$
\begin{equation} \label{eq:intJ}
J(Q^2, z) = \int_0^1 \! dx \, J_0\negthinspace \left(\negthinspace\zeta Q
\sqrt{\frac{1-x}{x}}\right) ,
\end{equation} we can write the AdS electromagnetic form-factor as
\begin{equation} 
F(Q^2)  =    R^3 \! \int_0^1 \! dx  \! \int \frac{dz}{z^3} \, 
J_0\!\left(\!z Q\sqrt{\frac{1-x}{x}}\right) \left \vert\Phi_{\pi^+}(z) \right\vert^2 .
\label{eq:AdSFx}
\end{equation}
Comparing with the expression for the electromagnetic form-factor in light-front 
QCD  (\ref{eq:PiFFb}) for arbitrary 
values of $Q$, we find the relation between the pion LFWF $\tilde \psi$
and the hadronic string mode $\Phi_\pi$~\cite{Brodsky:2007hb, Brodsky:2006uqa}
\begin{equation} \label{eq:PhipsiB} 
\left\vert \tilde\psi_{u \bar d/\pi}(x,\zeta)\right\vert^2 = 
\frac{R^3}{2 \pi} \, x(1-x)
\frac{\left\vert \Phi_\pi(\zeta)\right\vert^2}{\zeta^4}, 
\end{equation}
where we identify the transverse light-front variable $\zeta$, $0 \leq \zeta \leq \Lambda_{\rm QCD}$,
with the holographic variable $z$.

\subsection{Gravitational Form Factor}

The light-front expression for the helicity-conserving gravitational form factor in impact space is
(\ref{eq:MGFFbb})
\begin{multline} \label{eq:GFFb}
A(q^2) =  \sum_n  \prod_{j=1}^{n-1}\int d x_j d^2 \mbf{b}_{\perp j} \\
\times \sum_f x_f
\exp \! {\Bigl(i \mbf{q}_\perp \! \cdot \sum_{k=1}^{n-1} x_k \mbf{b}_{\perp k}\Bigr)} 
\left\vert \tilde \psi_n(x_j, \mbf{b}_{\perp j})\right\vert^2,
\end{multline}
which includes the contribution of each struck parton with longitudinal momentum fraction $x_f$ and
corresponds to a change of transverse momentum $x_j \mbf{q}$ for
each of the $j = 1, 2, \cdots, n-1$ spectators. 
For $n=2$, there are two terms which contribute to the $f$-sum in  (\ref{eq:GFFb}). 
Exchanging $x \leftrightarrow 1-x$ in the second integral we find
\begin{equation} \label{eq:GFF}
A_{\pi}(q^2) =   2 \! \int_0^1 \! x \, dx \int \! d^2 \mbf{b}_{\perp}  
 e^{i \mbf{q}_\perp \cdot  \mbf{b}_{\perp} (1-x)} 
\left\vert \tilde \psi_{q \bar q / \pi}\!  \left(x, \mbf{b}_{\perp } \right)\right\vert^2. 
\end{equation} 
Using the light-front wave function normalization
\begin{equation}
\int_0^1 \! dx  \int \! d^2 \mbf{b}_{\perp}  \big\vert \tilde \psi(x, \mbf{b}_\perp) \big\vert^2 = 1,
\end{equation}
it is simple to prove that if
$\psi$ is a symmetric function of $x$ and  $1-x$ the first $x$-moment
\begin{equation}
\int_0^1 \! x \,dx  \int \! d^2 \mbf{b}_{\perp}  
\big\vert \tilde \psi(x, \mbf{b}_\perp) \big\vert^2 = \frac{1}{2},
\end{equation}
and thus $A_\pi(q^2)$ satisfies the sum rule $A_\pi(0) = 1$.
Integrating (\ref{eq:GFF}) over angle we find
\begin{multline}  \label{eq:PiGFFb} 
A_\pi(Q^2) =  4 \! \pi \int_0^1 \frac{dx}{(1-x)} \\ \times \int \zeta d \zeta 
J_0 \! \left(\! \zeta q \sqrt{\frac{1-x}{x}}\right)  \vert\tilde\psi_{q \bar q / \pi}\! (x,\zeta)\vert^2,
\end{multline}
where $\zeta^2 =  x(1-x) \mathbf{b}_\perp^2$. 

We now consider the expression for the hadronic gravitational form factor in 
AdS space (\ref{eq:AdSA}) 
\begin{equation} 
A(Q^2)  =  R^3 \! \! \int \frac{dz}{z^3} \, H(Q^2, z) \left\vert\Phi_\pi(z) \right\vert^2,
\end{equation}
where $A(Q)$ is normalized to one at $Q\!=\!0$ and
$H(Q^2, z) = \half  Q^2 z^2  K_2(z Q)$
describes the propagation of the external graviton inside AdS space.
Using the integral representation of $H(Q^2,z)$ (\ref{eq:intHz})
\begin{equation} 
H(Q^2, z) =  2 \! \int_0^1\!  x \, dx \, J_0\!\left(\!z Q\sqrt{\frac{1-x}{x}}\right) ,
\end{equation}
the AdS gravitational form factor can be expressed as
\begin{equation} 
A(Q^2)  =  2  R^3 \! \int_0^1 \! x \, dx  \! \int \! \frac{dz}{z^3} \, 
J_0\!\left(\!z Q\sqrt{\frac{1-x}{x}}\right) \left \vert\Phi_\pi(z) \right\vert^2 .
\end{equation}
Comparing with the QCD  light-front gravitational form factor (\ref{eq:PiGFFb}) we find
($\zeta =z$)
\begin{equation} 
\left\vert \tilde\psi_{q \bar q / \pi} (x,\zeta)\right\vert^2 = 
\frac{R^3}{2 \pi} \, x(1-x)
\frac{\left\vert \Phi_\pi(\zeta)\right\vert^2}{\zeta^4}, 
\end{equation}
which is identical to the result (\ref{eq:PhipsiB}) obtained from the mapping of the pion electromagnetic transition amplitude.

\end{document}